\documentclass[aps,12pt,a4paper,onecolumn,showpacs,showkeys,titlepage,preprint]{revtex4}

\usepackage{times}
\usepackage[dvipdf]{graphicx}
\usepackage{amsmath}
\usepackage{amssymb}
\usepackage[latin1]{inputenc}
\newcommand{\be}{\begin{equation}}
\newcommand{\ee}{\end{equation}}

\begin{document}
\title{Solitons of two-component Bose-Einstein condensates\\ modulated in space and time}
\author{W.B. Cardoso$^a$, A.T. Avelar$^a$, D. Bazeia$^b$, M.S. Hussein$^{c,}$\footnote{Corresponding author.\\E-mail addresses:: bazeia@fisica.ufpb.br (D. Bazeia), hussein@if.usp.br (M.S. Hussein).}}
\affiliation{$^a$Instituto de F\'{\i}sica, Universidade Federal de Goi\'{a}s, 74.001-970, Goi\^{a}nia - Goi\'{a}s, Brazil\\
$^b$Departamento de F\'{\i}sica, Universidade Federal da Para\'{\i}ba, 58.051-970, Jo\~{a}o Pessoa - Para\'{\i}ba, Brazil\\
$^c$Instituto de F\'{i}sica, Universidade de S\~{a}o Paulo, 05314-970,
S\~{a}o Paulo, SP, Brazil}
%\author{A. T. Avelar}
%\affiliation{Instituto de F\'{\i}sica, Universidade Federal de Goi\'{a}s, 74.001-970, Goi\^{a}nia - Goi\'{a}s, Brazil}
%\author{D. Bazeia}
%\affiliation{Departamento de F\'{\i}sica, Universidade Federal da Para\'{\i}ba, 58.051-970, Jo\~{a}o Pessoa - Para\'{\i}ba, Brazil}
%\author{M. S. Hussein}
%\affiliation{Instituto de F\'{i}sica, Universidade de S\~{a}o Paulo, 05314-970,
%S\~{a}o Paulo, SP, Brazil}


\begin{abstract}
In this paper we present soliton solutions of two coupled nonlinear Schr\"{o}dinger equations modulated in space and time. The approach allows us to obtain solitons for a large
variety of solutions depending on the nonlinearity and potential
profiles. As examples we show three cases with soliton solutions: a
solution for the case of a potential changing from repulsive to attractive
behavior, and the other two solutions corresponding to localized and
delocalized nonlinearity terms, respectively.
\end{abstract}
\maketitle

Coupled nonlinear Schr\"{o}dinger equations (CNLSE) are very important
because they can be used to model a great variety of physical systems. Solitary waves in
these equations are often called vector solitons in the literature since they
are generically described by two-component wave functions \cite{Kivshar03,Malomed06}. One of the simplest vector solitons is known as
shape-preserving, self-localized solution of coupled nonlinear evolution
equations \cite{Kivshar03}. The CNLSE may also model beam propagation inside crystals
as well as water wave interactions, and in fiber communication system, such
equations have been shown to govern pulse propagation along orthogonal
polarization axes in nonlinear optical fibers and in
wavelength-division-multiplexed systems \cite{MenyukJOSAB98}.

In the case of Bose-Einstein condensates, \cite{String,Pethick}, a many-body description of  the effect of Feshbach resonances in the mean field limit, requires the use of two coupled Gross-Pitaevskii equations \cite{TTHK,TTCHK}. This results in a convenient way to account for the physics of the so-called super-chemistry, \cite{Drummond}.  In the case of one-dimensional (1D) Bose-Einstein condensates (BECs), the presence of {\it cigar-shaped} traps introduces tight confinement in two transverse directions, while leaving the condensate almost free along
the longitudinal axis. The realization of BEC of trapped atoms was experimentally achieved in \cite{AndersonSC95,BradleyPRL95,DavisPRL95}.
Next, dark solitons in BEC were formed with repulsive $^{87}$Rb atoms in \cite{BurgerPRL99,DenschlagSC00}, and bright solitons were
generated for attractive $^{7}$Li atoms in \cite{StreckerNAT02,KhaykovichSC02}. 
In the case of two-component condensates, the experimental generation has been achieved for different hyperfine states in rubidium
atoms in a magnetic trap \cite{MyattPRL97} and in sodium atoms in an optical trap \cite{Stamper-KurnPRL98}.

When treationg a set of CNLSE's, one usually deals with two or more coupled NLSE's with constant coefficients. More recently, however, in
Refs. \cite{SerkinPRL00,SerkinPRL07,Belmonte-BeitiaPRL08} the authors have shown how to deal with NLSE with varying cubic
coefficient, and in \cite{AvelarPRE09} one has extended the procedure to the case of cubic and quintic nonlinearities.
With this motivation on mind, our goal in the present work is to deal with soliton solutions of the CNLSE where the
potentials and nonlinearities are modulated in space and time. The procedure consists in choosing an {\it Ansatz} which
transforms two CNLSE's into a pair of CNLSE's with constant coefficients, certainly much easier to solve. With this procedure we can
then obtain analytic solutions for some specific choices of parameters, leading us to interesting localized coupled
soliton solutions of the bright and/or dark type. We illustrate the procedure with several distinct possibilities:
firstly, we consider the system with attractive potentials and localized nonlinearities. Next, we consider nonlinearities
delocalized and then we make the potential to change periodically in time from attractive to repulsive behavior.

The model we start with is described by two CNLSE's, with interactions up to third-order. These equations are given by
\begin{equation}
i\frac{\partial\psi_1}{\partial t}=-\frac{\partial^{2}\psi_1}{\partial x^{2}}
+v_{1}(x,t)\psi_1 +\sum_{k=1}^2( g_{1k}(x,t)|\psi_k|^{2})\psi_1,\label{NLS1}
\end{equation}
\begin{equation}
i\frac{\partial \psi_2}{\partial t}=-\frac{\partial^{2}\psi_2}{\partial x^{2}}
+v_{2}(x,t)\psi_2+\sum_{k=1}^2( g_{2k}(x,t)|\psi_k|^{2})\psi_2,  \label{NLS2}
\end{equation}
where $\psi_k=\psi_k(x,t)$, the functions, $v_{k}(x, t)$, are the trapping potentials, and the $g_{ij}, (i,j=1,2)$ describe the strength of the cubic nonlinearities. Note that we are using
standard notation, with both the fields and coordinates dimensionless. Note also that we are supposing that our model describe
1D BEC, and then it should engender {\it cigar-shaped} traps to induce tight confinement in two transverse directions, leaving
the condensate almost free along the $x$ axis. In the super-chemistry model of \cite{TTCHK,Drummond}, the above equations are amended by a term which goes as
$\sqrt{2} \alpha \psi_1\psi_2$ in Eq. (\ref{NLS1}) and with a term of $\frac{\alpha}{\sqrt{2}}\psi_{1}^2$ in Eq. (\ref{NLS2}). These, number-nonconserving, terms arise from treating
the Feshbach resonance as a coupling term between atoms and molecules. 

It is interesting to comment here that the elimination of, say, $\psi_2$ in favor of an effective equation for $\psi_1$, would necessarily end up with a GP equation with depletion effect. This could appear in the form of complex nonlinearity. We leave such study for a future work.

In the following we seek to reduce the above two equations into the following pair of equations
\begin{equation}
\mu_{1}A_1=-\frac{\partial^{2}A_1}{\partial\zeta^{2}}+\sum_{k=1}^2\left(
G_{1k}|A_k|^{2}\right) A_1,  \label{STA}
\end{equation}%
\begin{equation}
\mu_{2}A_2=-\frac{\partial^{2}A_2}{\partial\zeta^{2}}+\sum_{k=1}^2\left(
G_{2k}|A_k|^{2}\right) A_2,  \label{STB}
\end{equation}
with $\mu_i$ and the $G_{ij}$ being constants. We achieve this with the use of the two{\it Anzatse}:
\begin{equation}
\psi_1(x,t)=\rho(x,t)e^{i\eta(x,t)}A_1(\zeta(x,t)),  \label{ST1}
\end{equation}
\begin{equation}
\psi_2(x,t)=\rho(x,t)e^{i\eta(x,t)}A_2(\zeta(x,t)).  \label{ST2}
\end{equation}
However, we now have to have
\begin{equation}
\rho\frac{\partial\rho}{\partial t}+\frac{\partial}{\partial x}\left(
\rho^{2}\frac{\partial\eta}{\partial x}\right) =0,  \label{e7}
\end{equation}
\begin{equation}
\frac{\partial\zeta}{\partial t}+2\frac{\partial\eta}{\partial x}\frac{\partial\zeta}{\partial x}=0,\label{e8}
\end{equation}
\begin{equation}
\frac{\partial}{\partial x}\left( \rho^{2}\frac{\partial\zeta}{\partial x}\right)=0.  \label{e9}
\end{equation}
These equations control the functions $\rho(x,t)$, $\eta(x,t)$ and $\zeta(x,t)$, and the two potentials are given by
\begin{equation}
v_{i}(x,t)=\frac{1}{\rho }\frac{\partial ^{2}\rho }{\partial x^{2}}-\frac{%
\partial \eta }{\partial t}-\left( \frac{\partial \eta }{\partial x}\right)
^{2}-\mu _{i}\left( \frac{\partial \zeta }{\partial x}\right) ^{2}.
\label{pot}
\end{equation}

We take $\zeta(x,t)=F(\xi (x,t))$, with $\xi (x,t)=x/\chi(t)$, which has to obey
the Eqs.~(\ref{e7}), (\ref{e8}) and (\ref{e9}). After some algebraic calculations we get that
\begin{equation}
g_{jk}=\frac{G_{jk}}{\rho ^{6}\chi ^{4}}.  \label{g}
\end{equation}
We use Eq.~(\ref{e9}) to obtain 
\begin{equation}
\rho(x,t)=1/\sqrt{\chi F^{\prime }(\xi (x,t))}.  \label{rho}
\end{equation} 
Also, using Eq. (\ref{e8}) we can write 
\begin{equation}
\eta (x,t)=\frac{1}{4\chi }\frac{\partial \chi }{\partial t}x^{2}+a,
\label{eta}
\end{equation}%
where $a=a(t)$ is a function of time. 

The above results can be used to find finite energy solutions for both $\psi_1$ and $\psi_2$, with appropriate choices
of $F(\xi)$, according to Eq.~(\ref{rho}). We illustrate these results with some applications of current interest
to BEC.

%%%%%%%%%%%%%%%%%%%%%%%%%%%%%%%%%%%%%%%%%%%%%%%%%%%%%%%
\textit{Example \# 1.} --  The first example is a extension of the model discussed in \cite{Belmonte-BeitiaPRL08},
in the case of nonlinearities given by
\be
g_{jk}=G_{jk}\chi ^{-1}\exp (-3\xi ^{2}),
\ee
with $\chi=\chi(t)>0$. We consider the simpler case, with $\mu_{1}=\mu_2=0$. Here, we get that $v_1(x,t)=v_2(x,t)=v(x,t)$, with $v(x,t)$ being an attractive potential.
We use Eq.~(\ref{g}) to get $\rho(x,t)=\exp (\xi ^{2}/2)/\sqrt{\chi }$ and the potential is now given by
\begin{equation}
v(x,t)=f(t)\,x^{2}+h(t),\label{vi}
\end{equation}%
with
\begin{equation}
f(t)=\frac{1}{\chi ^{4}}-\frac{1}{4\chi }\frac{d^{2}\chi }{dt^{2}},
\label{f11}
\end{equation}%
\begin{equation}
h(t)=\frac{1}{\chi ^{2}}-\frac{da}{dt}.  \label{f21}
\end{equation}%
With the choice $a=\int \chi ^{-2}dt$, we can write $h(t)=0$. Thus, the potentials acquire the simple
form $v(x,t)=f(t)\,x^{2}$. Here we choose $G_{11}=-1/2$, $G_{12}=-1,$ $G_{21}=-1/2$, $G_{22}=-1$,
which we use in Eqs.~(\ref{STA}-\ref{STB}). In this case, each one of the two solutions, $\psi_1$ and $\psi_2$, describes a similar group of atoms with attractive behavior, as is the group of atoms investigated in \cite{Belmonte-BeitiaPRL08}. With the above choice of parameters we obtain the solutions 
\begin{equation}
A_1(x,t)=\frac{A_{10}\text{sn}(A_{10}\zeta ,1/\sqrt{2})}{\sqrt{2}\text{dn}%
(A_{10}\zeta ,1/\sqrt{2})},
\end{equation}%
\begin{equation}
A_2(x,t)=\frac{A_{20}\text{sn}(A_{20}\zeta ,1/\sqrt{2})}{2\text{dn}(A_{20}\zeta
,1/\sqrt{2})},
\end{equation}%
where $A_{10}=A_{20}=A_0=2n$K$(1/\sqrt{2})/\sqrt{\pi}$, for $n=1,2,...$ and K$(1/\sqrt{2})$ being the elliptic integral. The value of $A_0$ is obtained through
appropriate boundary conditions; see \cite{Belmonte-BeitiaPRL08}.

We use Eq.~(\ref{rho}) to get $\zeta=\frac{\sqrt{\pi}}{2}\left[1+\text{erf}\left(\xi\right)\right]$. Thus, the pair of coupled solutions are given by 
\begin{equation}
\psi_1(x,t)=e^{i\eta}\frac{A_{0}\exp (\xi^{2}/2)}{\sqrt{2}\sqrt{\chi }}
\frac{\text{sn}(A_{0}\zeta ,1/\sqrt{2})}{\text{dn}(A_{0}\zeta ,1/\sqrt{2})},
\label{sol11}
\end{equation}
\begin{equation}
\psi_2(x,t)=e^{i\eta }\frac{A_{0}\exp(\xi ^{2}/2)}{2\sqrt{\chi }}\frac{\text{
sn}(A_{0}\zeta ,1/\sqrt{2})}{\text{dn}(A_{0}\zeta ,1/\sqrt{2})},
\label{sol21}
\end{equation}
with $\eta$ being a real function obeying Eq.~(\ref{eta}).

Next, we use Eq.~(\ref{f11}) to obtain the function $\chi(t)$. As two distinct examples, we
consider the cases $f=1$ and $f=1+\epsilon\cos(\omega_{0}t)$, to describe periodic and quasiperiodic 
possibilities, respectively. The first case, $f=1$, leads us to $\chi=[1+15\cos^{2}(2t)]^{1/2}/2$.
The second case, $f=1+\epsilon\cos(\omega_{0}t)$, is a little more involved; it leads us to $\chi$
which is given as a combination of the two solutions of the Mathieu equation
$\frac{d^{2}z}{dt^{2}}+4f(t)z=0$. It has the form $\chi=(2z_{1}^{2}+2z_{2}^{2}/W^{2})^{1/2}$,
where $z_{1,2}$ represent the two linearly independent solutions of the Mathieu equation, with $W$ being
the corresponding Wronskian.

\begin{figure}
\includegraphics[{width=4cm}]{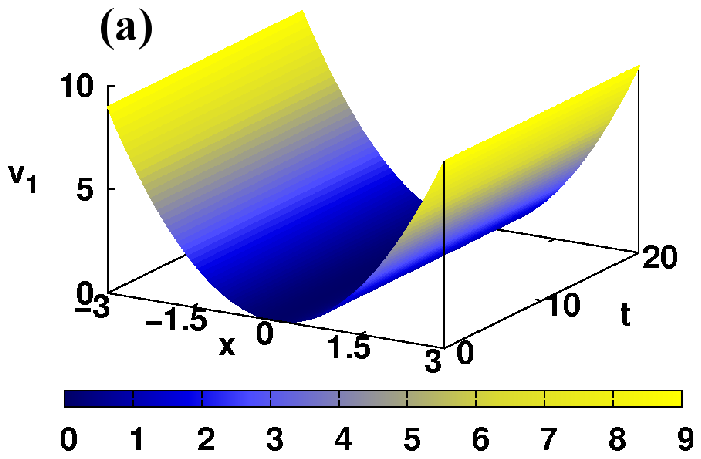} %
\includegraphics[{width=4cm}]{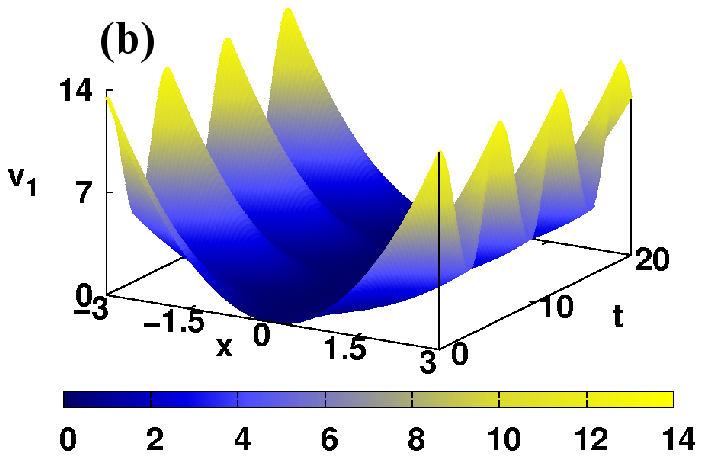}
\caption{Plot of the potential $v_{1}$ for the cases for (a) periodic and (b)
quasiperiodic, for $\protect\epsilon=0.5$ e $\protect\omega _{0}=1$.}
\label{potf1}
\end{figure}

\begin{figure}
\includegraphics[{width=4cm}]{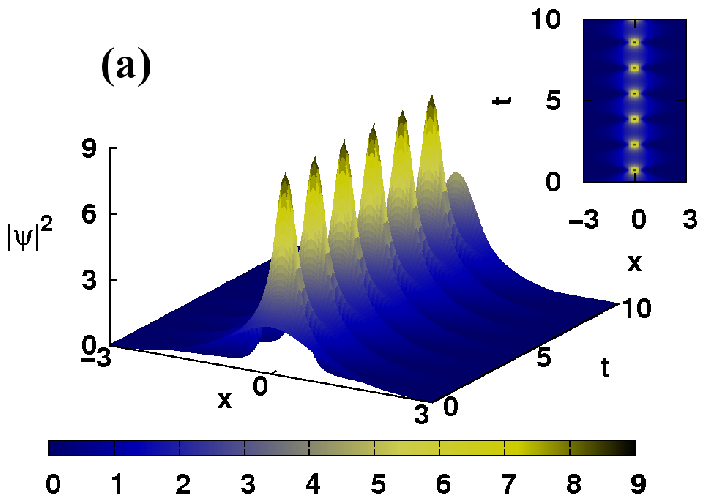}
\includegraphics[{width=4cm}]{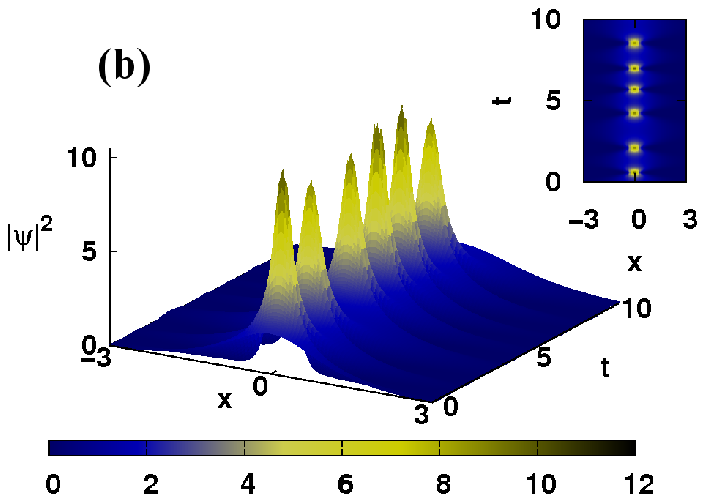}
\caption{Plot of $|\protect\psi_1|^{2}$ (Eq.~(\protect\ref{sol11})) for the two cases
(a) periodic and (b) quasiperiodic, for $n=1$, $\protect\epsilon=0.5$ and $\protect\omega _{0}=1$.}
\label{soly1}
\end{figure}

In Fig.~$\ref{potf1}$ we show the the potential $v(x,t)$ for the two cases, (a) periodic, and (b) quasiperiodic.
The solution (\ref{sol11}) is plotted in Fig.~\ref{soly1}, for the periodic and quasiperiodic cases, respectively.
The initial data for the Mathieu equation are $z(0)=\sqrt{2}$ and ${dz}/{dt}(0)=0$. The results for $\psi_2$ are similar
and  we do not show them.

To verify stability of the solutions, we use the split-step method, through finite difference to investigate the propagation
of the initial states $\psi_1(x,0)$ and $\psi_2(x,0)$ given by Eqs.~(\ref{sol11}) and (\ref{sol21}), at $t=0$, with the inclusion
of a small aleatory perturbation of $3\%$ on the profile of the initial states. With this procedure, we could verify that the solutions
(\ref{sol11}) and (\ref{sol21}) are both stable against such aleatory perturbations.

%%%%%%%%%%%%%%%%%%%%%%%%%%%%%%%%%%%%%%%%%%%%%%%%%%%%%%%
\textit{Example \# 2} -- In this case, we consider an attractive
potential but we allow the nonlinearity to change, as we did  
in Ref.~\cite{AvelarPRE09}. We choose the nonlinearity terms to be, 
\be
g_{jk}=G_{jk}\chi^{-1}\exp(\xi^{2}/\gamma^{2})
\ee
where $\gamma$ gives the width of the nonlinearities. With this, we can use Eqs.~(\ref{g}) and (\ref{rho}) to find
$\rho(x,t)=\exp(-\xi ^{2}/6\gamma^{2})/\sqrt{\chi}$ and $\zeta(\xi)=\int\exp(-\xi^{2}/3\gamma^{2})d\xi$, respectively.
The potentials are now given by
\begin{equation}
v_{j}=f(t)x^{2}+h(t)-\mu _{j}\exp(-2\xi^{2}/3\gamma^{2}),
\label{vii}
\end{equation}
with $f(t)$ e $h(t)$ given by Eqs.~(\ref{f11}) and (\ref{f21}). Here we use $a=\int\chi^{-2}dt$, which
leads us to $h(t)=0$. In this case, we can obtain solutions of the form 
\begin{equation}
A_1(x,t)={\rm sech}(\zeta(x,t)),
\end{equation}
\begin{equation}
A_2(x,t)=\frac{1}{\sqrt{2}}{\rm sech}(\zeta (x,t)),
\end{equation}%
with the choice $\mu _{1}=$ $\mu _{2}=G_{11}=-1$, $G_{12}=G_{21}=-2$,
and $G_{22}=0$, in the Eqs.~(\ref{STA}-\ref{STB}). In this case, we
note that $\psi_1$ describes a group of atoms similar to the group
considered in \cite{AvelarPRE09},(but without quintic interaction)
which interacts with the other group of atoms described by
$\psi_2$. This last group, however, does not self-interact. Here, the general solution has the form
\begin{equation}
\psi_1(x,t)=e^{i\eta}\chi^{-1/2}\exp(-\xi^{2}/6\gamma^{2}){\rm sech}(\zeta(x,t)),  \label{sol12}
\end{equation}
\begin{equation}
\psi_2(x,t)=e^{i\eta }(2\chi)^{-1/2}\exp(-\xi^{2}/6\gamma^{2}){\rm sech}(\zeta(x,t)).  \label{sol22}
\end{equation}

\begin{figure}
\includegraphics[{width=4cm}]{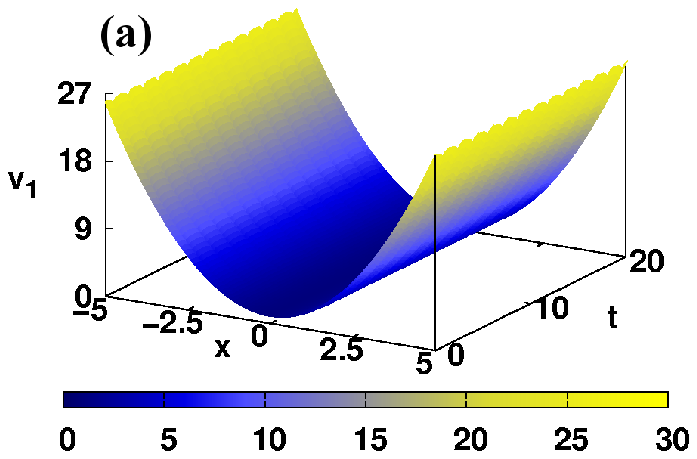} 
\includegraphics[{width=4cm}]{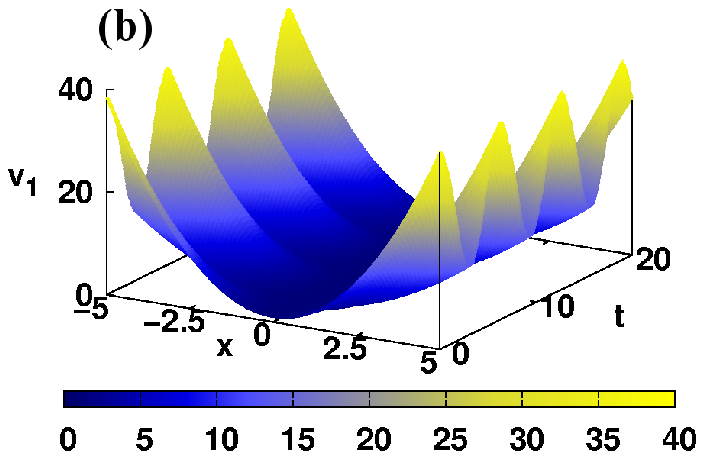}
\caption{Plot of the potential $v(x,t)$ (Eq.~(\protect\ref{vii})) for the cases (a) periodic and (b) quasiperiodic, for $\protect\gamma =6$,
$\protect\epsilon=0.5$ and $\protect\omega _{0}=1$.}
\label{potf2}
\end{figure}

\begin{figure}
\includegraphics[{width=4cm}]{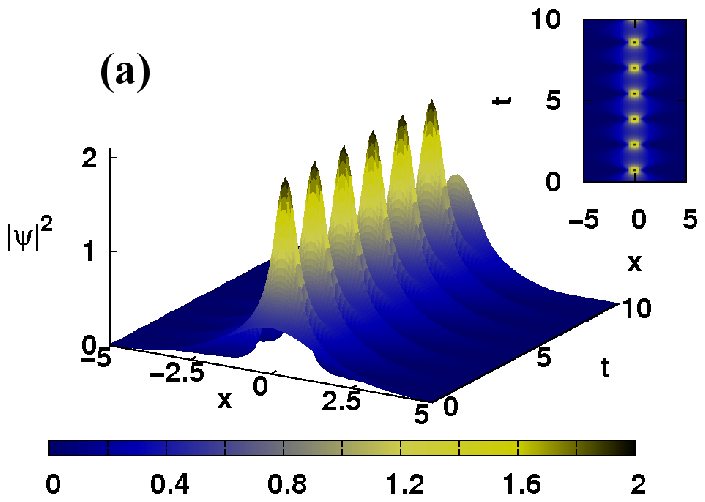} 
\includegraphics[{width=4cm}]{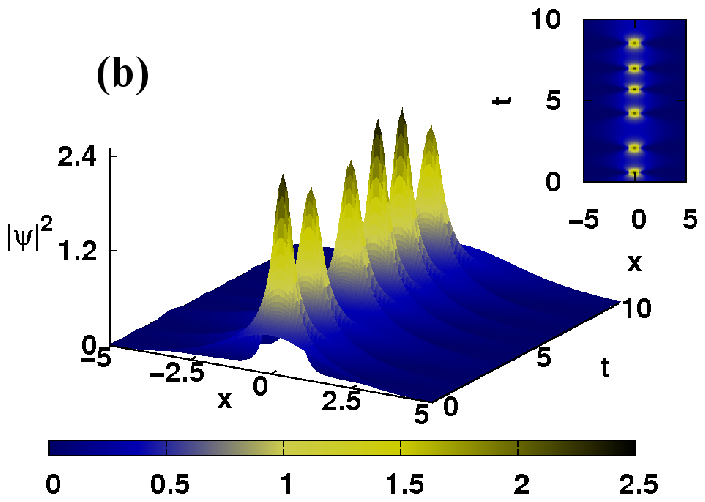}
\caption{Plot of $|\protect\psi_1|^{2}$ (Eq.~(\protect\ref{sol12})) for the (a) periodic and (b) quasiperiodic, for $\protect\gamma=6$,
$\protect\epsilon=0.5$ and $\protect\omega _{0}=1$.}
\label{soly2}
\end{figure}

As before, we see that the temporal evolution of the solutions are directly related to $\chi^{-1/2}$, which can be obtained from Eq.~(\ref{f11}).
For $f=1$ and for $f=1+\epsilon\cos(\omega_{0}t)$, we can get periodic and quasiperiodic solutions, respectively, as show before in the former example.
In Fig.~\ref{potf2} we show the potential for the case (a) periodic and (b) quasiperiodic, respectively. The solution (\ref{sol12}) is plotted in
Fig.~\ref{soly2}, for the,(a) periodic, and (b) quasiperiodic, cases, respectively. As before, we do not depict the case $\psi_2$, since it is similar
to $\psi_1$.

We have studied numerically the propagation of the initial states $\psi_1(x,0)$ e $\psi_2(x,0)$, given by Eqs.~(\ref{sol12}) and (\ref{sol22}), at $t=0$,
after introducing aleatory perturbation of $3\%$ in the profile of the initial states. With this, we could verify that the solutions (\ref{sol12}) and
(\ref{sol22}) are stable against such aleatory perturbations.  

%%%%%%%%%%%%%%%%%%%%%%%%%%%%%%%%%%%%%%%%%%%%%%%%%%%%%%%
\textit{Example \# 3} -- Next, we consider the case with $\mu _1=\mu_2=G_{11}=0$, $G_{12}=-2$, $G_{21}=2$, and $G_{22}=-1$. We use this in
Eqs.~(\ref{STA}-\ref{STB}). This leads to two interesting analytical solutions: one
field leading to a bright-like solution, and the other giving rise to a dark-like solution.
They are given explicitly by 
\begin{equation}
A_1(x,t)=\frac{1}{\sqrt{2}}\tanh (\zeta (x,t)),\label{sol1}
\end{equation}
\begin{equation}
A_2(x,t)=\rm{sech}(\zeta(x,t)).\label{sol2}
\end{equation}
We take
\be
g_{jk}=G_{jk}\chi^{-1}(1+\lambda\exp(-\xi^{2}))^{3},
\ee
with $\chi(t)>0$. We use Eq.~(\ref{g}) to get
\be
\rho(x,t)=\frac{1}{\left(\chi\left(1+\lambda\exp(-\xi^{2})\right)\right)^{1/2}}
\ee
Now, the potentials are given by
\begin{equation}
v_{j}(x,t)=s_{1}(t)x^{2}+s_{2}(x,t),\label{viii}
\end{equation}
with
\begin{equation}
s_{1}=-\frac{1}{4\chi }\frac{d^{2}\chi }{dt^{2}},\label{f1}
\end{equation}
\begin{eqnarray}
s_{2}=\frac{\lambda\exp(-\xi^{2})}{\chi^{2}[1+\lambda\exp(-\xi^{2})]}\left(1+\frac{[\lambda\exp(-\xi^{2})-2]}{\chi^{2}[1+\lambda\exp(-\xi^{2})]}x^{2}\right).
\end{eqnarray}
This choice of nonlinearity gives $\zeta(x,t)=\xi+\sqrt{\pi}\lambda $erf$(\xi )/2$. With this, the solutions are given by 
\begin{equation}
\psi_1(x,t)=\frac{e^{i\eta}}{\sqrt{2}}\left[\chi\left(1+\lambda e^{-\xi^{2}}\right)\right]^{-1/2}\tanh(\zeta),  \label{sol13}
\end{equation}
\begin{equation}
\psi_2(x,t)=e^{i\eta}\left[\chi\left(1+\lambda e^{-\xi^{2}}\right)\right]^{-1/2}\rm{sech}(\zeta),  \label{sol23}
\end{equation}
with $\eta $ real, given by Eq.~(\ref{eta}), and $\zeta(x,t)$ as shown above. Due to the several kinds of modulations of the nonlinearities, we can choose $\chi(t)$ in several distinct ways.
For instance, we may choose $\chi(t)=1+\alpha\sin(t)+\beta\sin(\sqrt{2}t)$, with $\alpha$ and $\beta $ being constants related to the periodic or quasiperiodic choice of $\chi$.

In Fig.~(\ref{pot1}), we plot the potential (\ref{viii}) for the (a) periodic and (b) quasiperiodic cases. Since these potentials have some structure at small distances, in Fig.~(\ref{pot2}) we show the same potentials, but now for small $x$. The parameters were chosen to be $\alpha=0.1$ and $\beta=0$ for the periodic solution, and $\alpha=\beta =0.1$ for the quasiperiodic solution, with $\lambda=0.5$. Despite the presence of structures at small distances, the main importance of the potentials are their long distance behavior, which allows the presence of localized excitations, even though the potential changes from attractive to repulsive behavior periodically. We get to this result after changing $\lambda\to-\lambda$, which significantly alters the small distance behavior of the potential, but preserving its long distance behavior; even in this case, the system supports similar soliton solutions, showing that the small distance behavior of the potential does not affect the presence of localized solutions.  This is clearly seen in Figs. ~(\ref{s1}) and (\ref{s3}) where we show the corresponding solutions  $|\psi_1|^2$ and $|\psi_2|^2$, for the periodic and quasiperiodic choices of $\chi(t)$ considered above.  

Due to the presence of a dark-like solution, which does not vanish in the limit $x\rightarrow\pm\infty$, the numerical method used in the former two examples fails to work in such a case.  This, therefore, requires seeking alternative numerical algorithm to circumvent this problem. We are currently investigating this issue.

In summary, in this paper we have presented  soliton solutions for two coupled nonlinear Schr\"{o}dinger equations modulated in the space and time. This model is robust
in the sense of presenting a vast quantity of nontrivial solutions in the systems with modulations in space and time of the traps and nonlinearities. We
have considered three different examples of potentials and nonlinearities. All these solutions are of great interest in several areas, such that, in coupled
Bose-Einstein condensates and communications in optical fibers. Our work can potentially motivate future studies and help guide possible experimental work in superchemistry and fluid dynamics in general.

\section{Acknowledgments}
This work was supported by the CAPES, CNPq, FAPESP, FUNAPE-GO and the Instituto Nacional de Ci\^{e}ncia e Tecnologia da Informa\c{c}\~{a}o Qu\^{a}ntica-MCT.

%%%%%%%%%%%%%%%%%%%%%%%%%%%%%%%%%%%%%%%%%%%%%%%%%%%%%%%%

%%%%%%%%%%%%%%%%%%%%%%%%%%%%%%%%%%%%
\newpage

\textbf{FIGURE CAPTIONS:}

\phantom{a}

FIG. 1: Plot of the potential $v_{1}$ for the cases for (a) periodic and (b)
quasiperiodic, for $\protect\epsilon=0.5$ e $\protect\omega _{0}=1$.

\phantom{a}

FIG. 2: Plot of $|\protect\psi_1|^{2}$ (Eq.~(\protect\ref{sol11})) for the two cases
(a) periodic and (b) quasiperiodic, for $n=1$, $\protect\epsilon=0.5$ and $\protect\omega _{0}=1$.

\phantom{a}

FIG. 3: Plot of the potential $v(x,t)$ (Eq.~(\protect\ref{vii})) for the cases (a) periodic and (b) quasiperiodic, for $\protect\gamma =6$,
$\protect\epsilon=0.5$ and $\protect\omega _{0}=1$.

\phantom{a}

FIG. 4: Plot of $|\protect\psi_1|^{2}$ (Eq.~(\protect\ref{sol12})) for the (a) periodic and (b) quasiperiodic, for $\protect\gamma=6$,
$\protect\epsilon=0.5$ and $\protect\omega _{0}=1$.

\phantom{a}

FIG. 5: Plot of the potential $v(x,t)$ (Eq.~(\protect\ref{pot})) for the cases (a) periodic and (b) quasiperiodic.

\phantom{a}

\begin{figure}
\includegraphics[{width=4cm}]{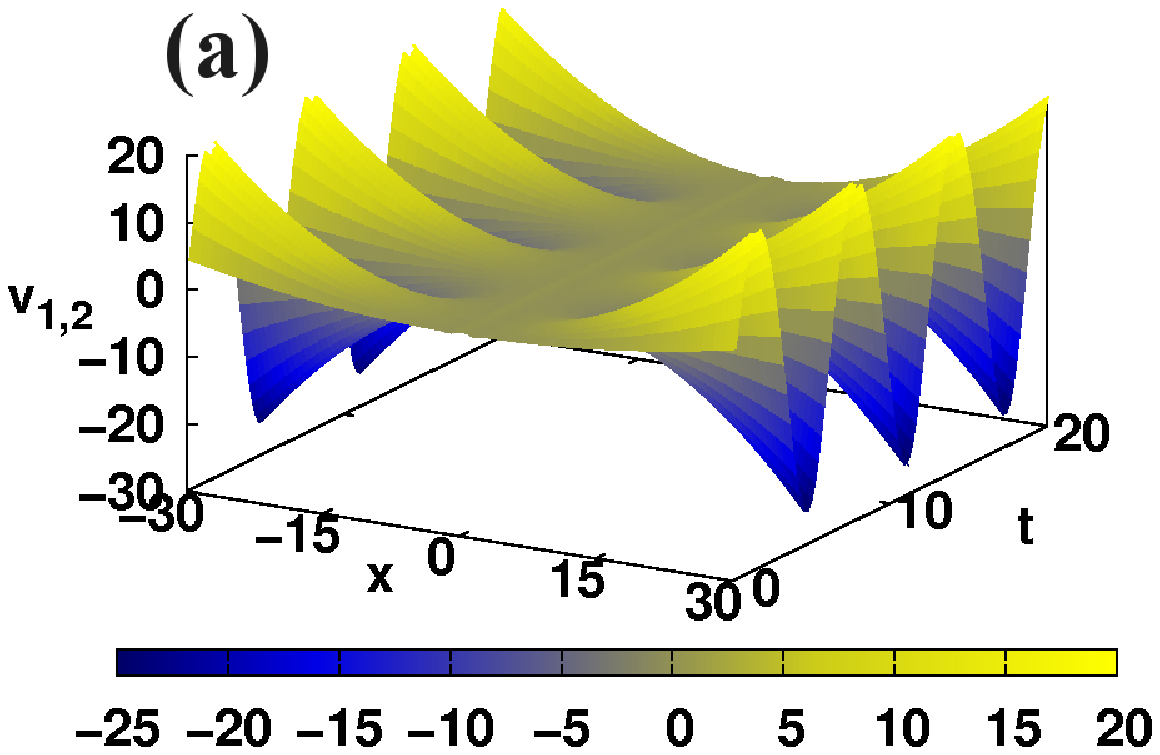} %
\includegraphics[{width=4cm}]{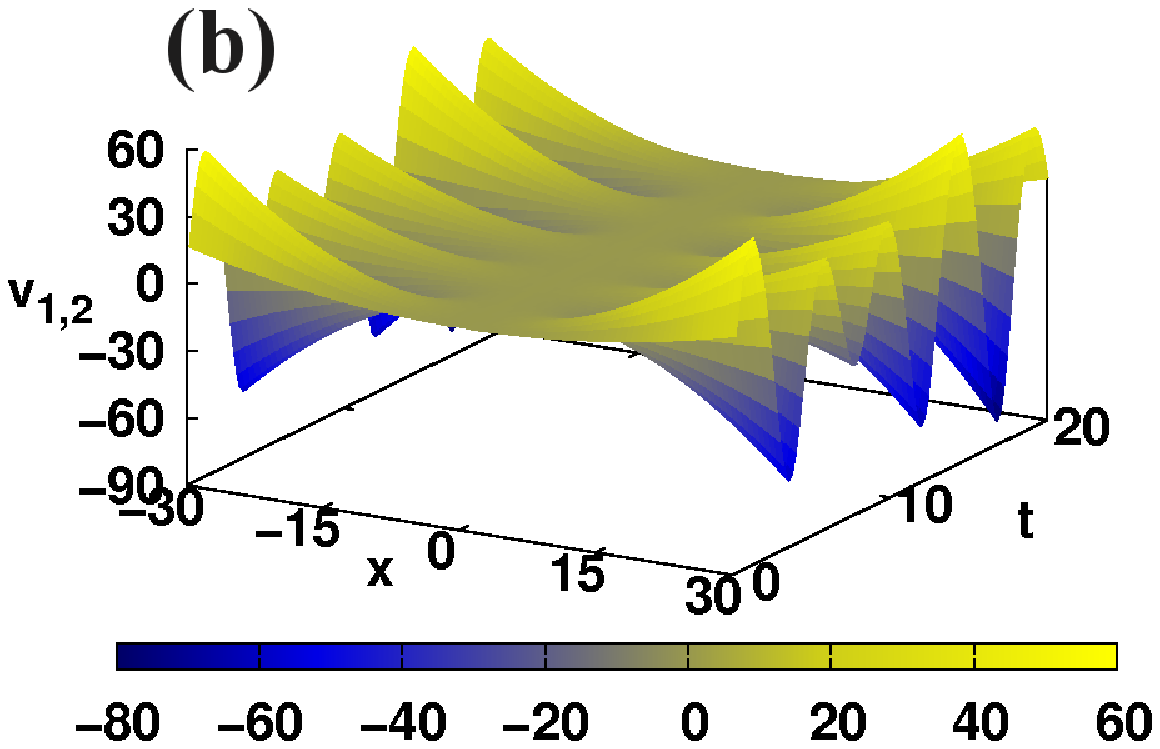}
\caption{Plot of the potential $v(x,t)$ (Eq.~(\protect\ref{pot})) for the cases (a) periodic and (b) quasiperiodic.}
\label{pot1}
\end{figure}

\begin{figure}
\includegraphics[{width=4cm}]{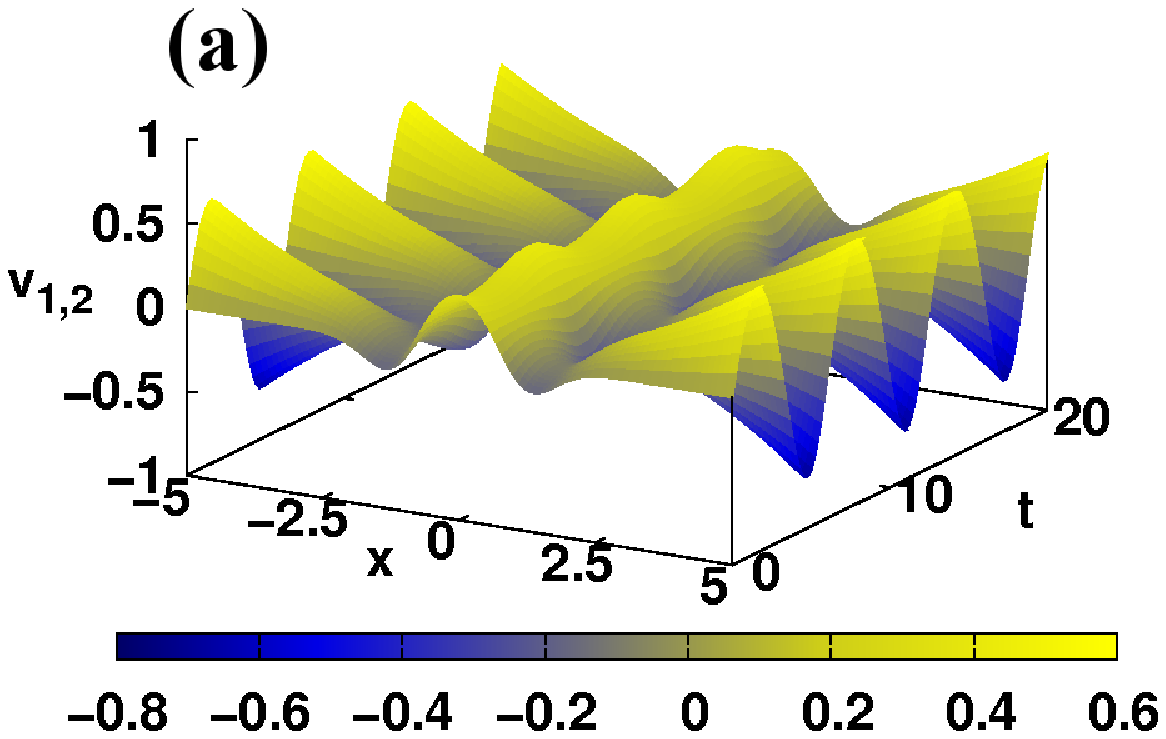} %
\includegraphics[{width=4cm}]{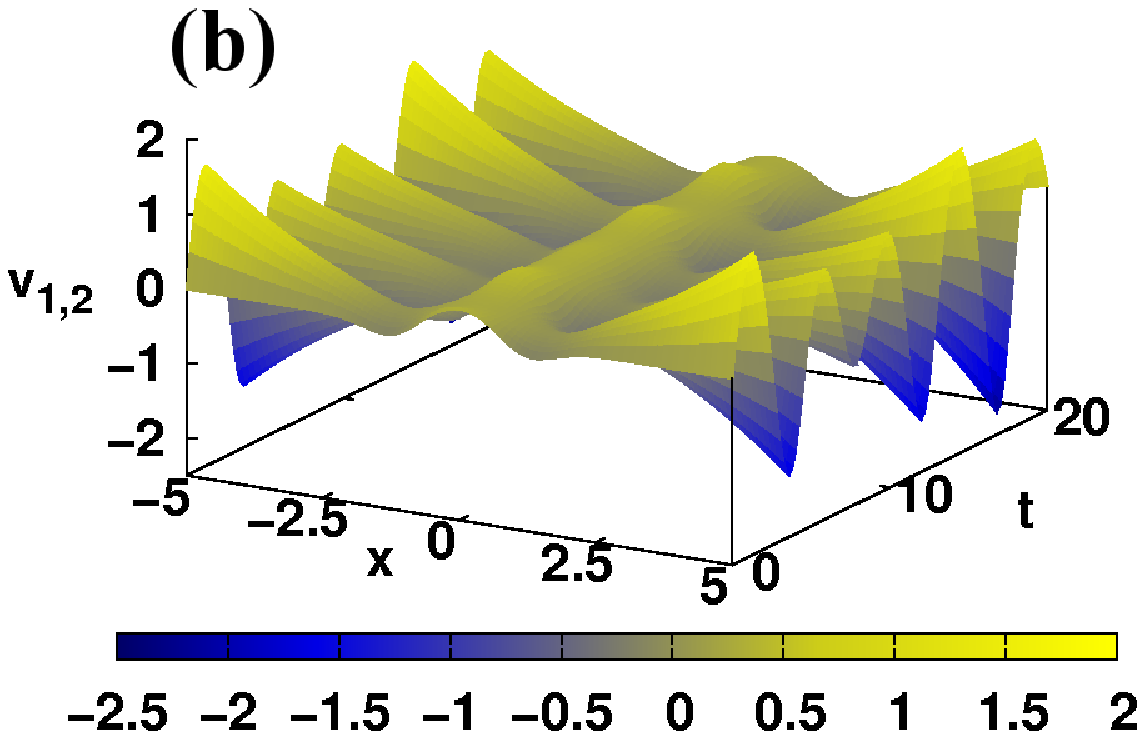}
\caption{Plot of the potentials show in Fig.~\protect\ref{pot1} for $x$ small. In (a) we display the case periodic, and in (b) the quasiperiodic case is shown.}
\label{pot2}
\end{figure}

\begin{figure}
\includegraphics[{width=4.2cm}]{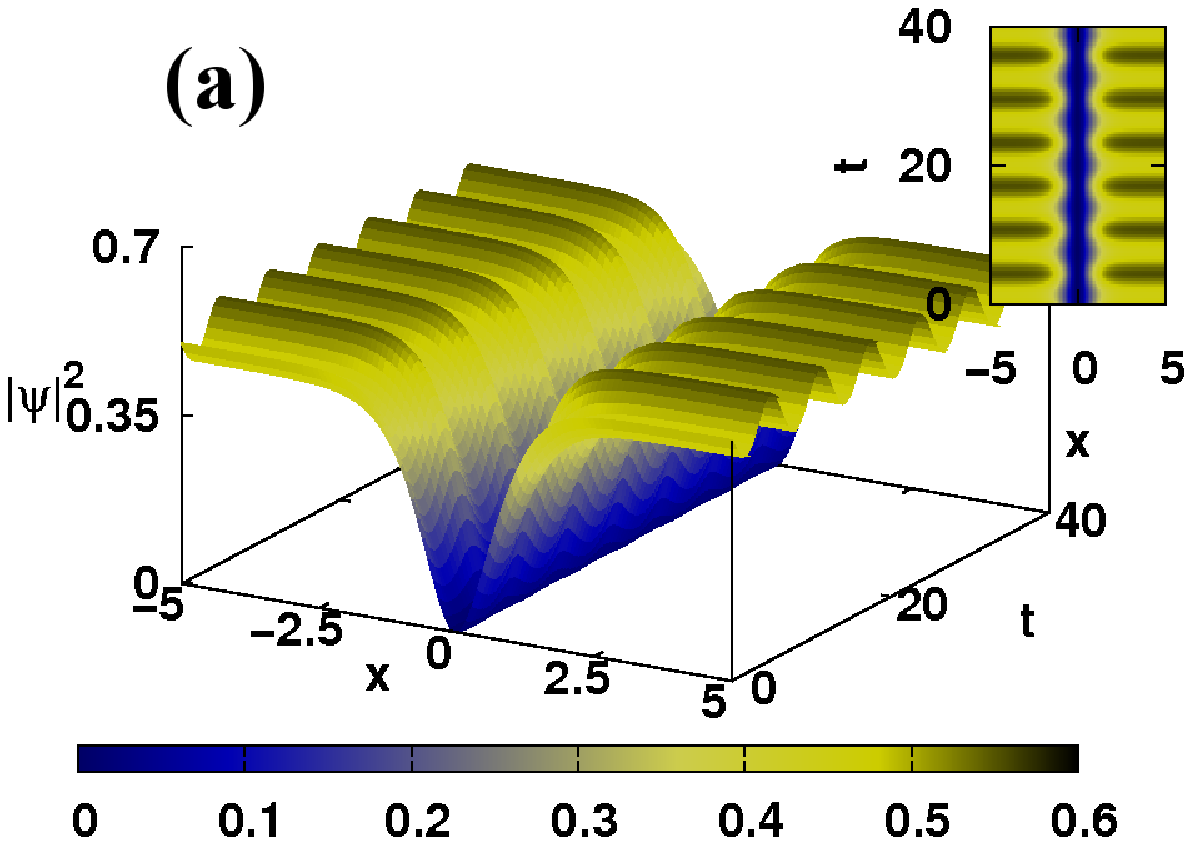} 
\includegraphics[{width=4.2cm}]{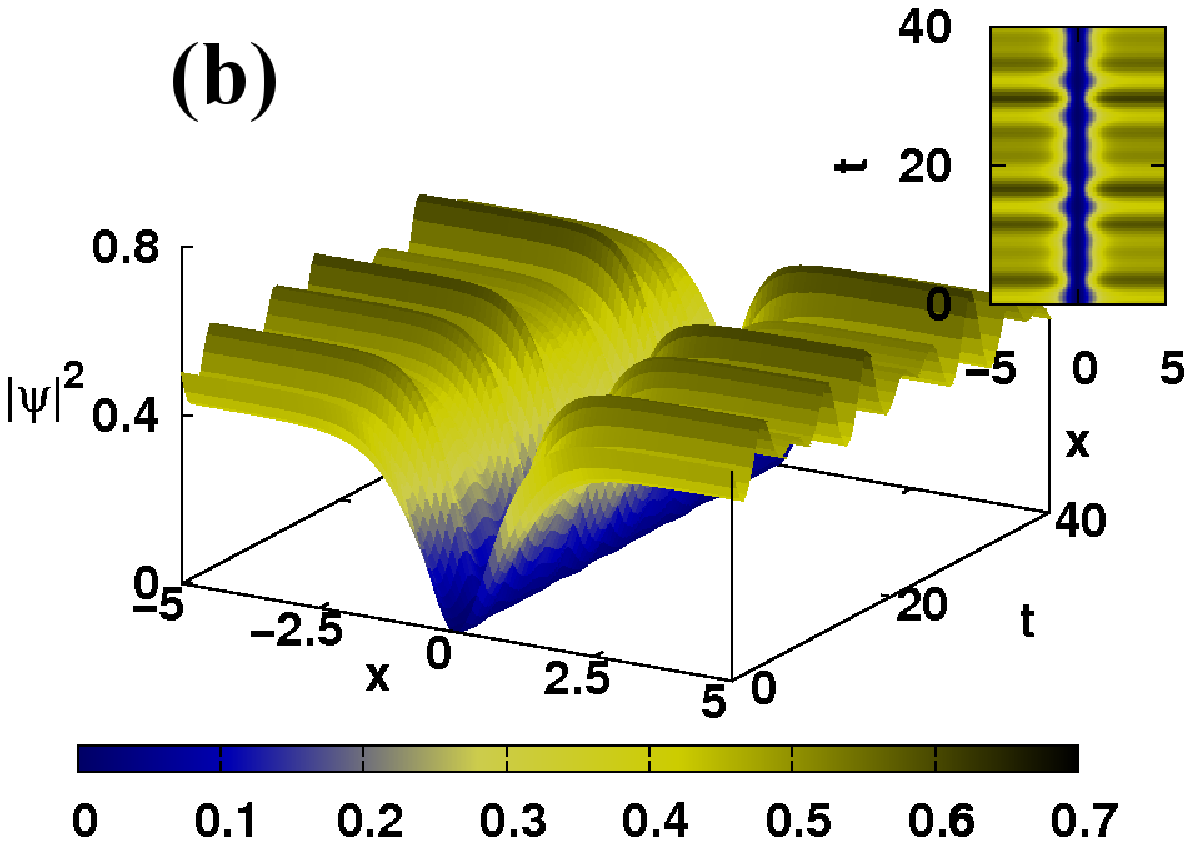}
\caption{Plot of $|\protect\psi_1|^{2}$ (Eq.~(\protect\ref{sol13})) for the cases (a) periodic and (b) quasiperiodic.}
\label{s1}
\end{figure}

\begin{figure}
\includegraphics[{width=4.2cm}]{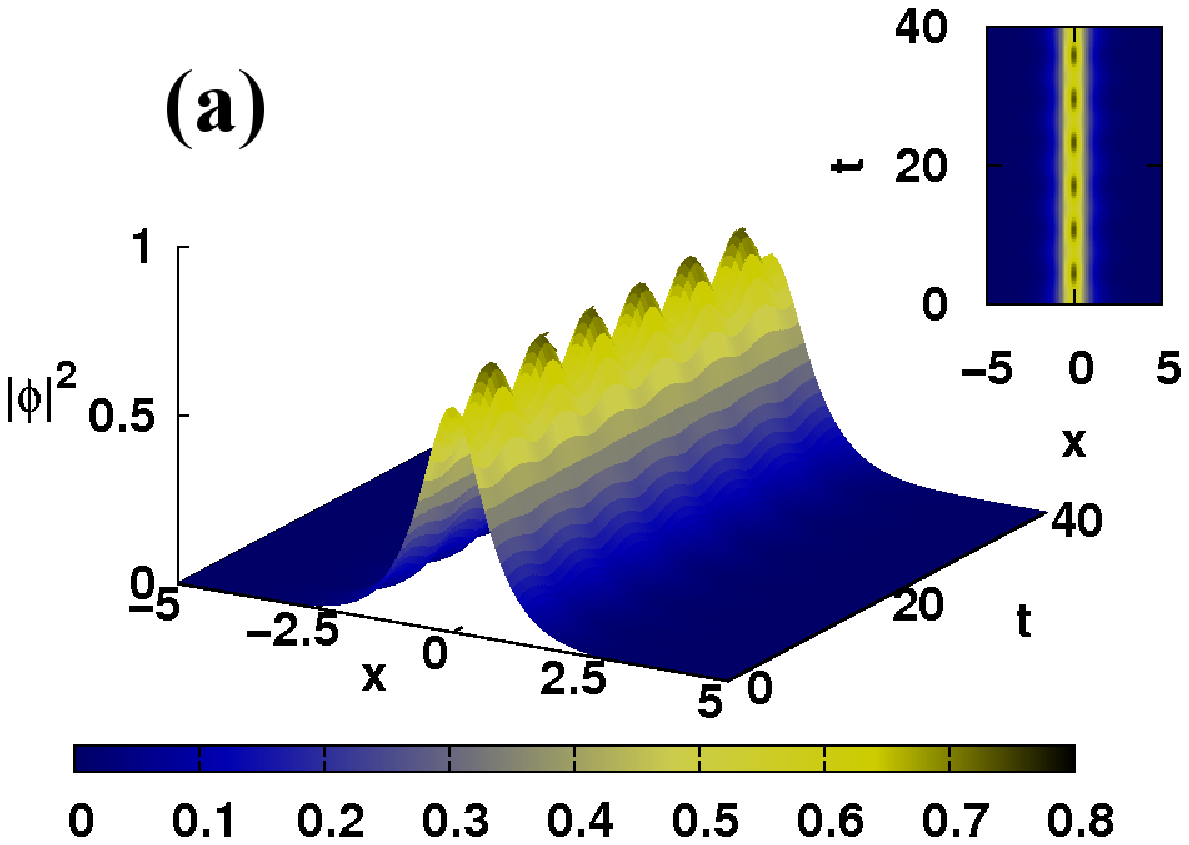} %
\includegraphics[{width=4.2cm}]{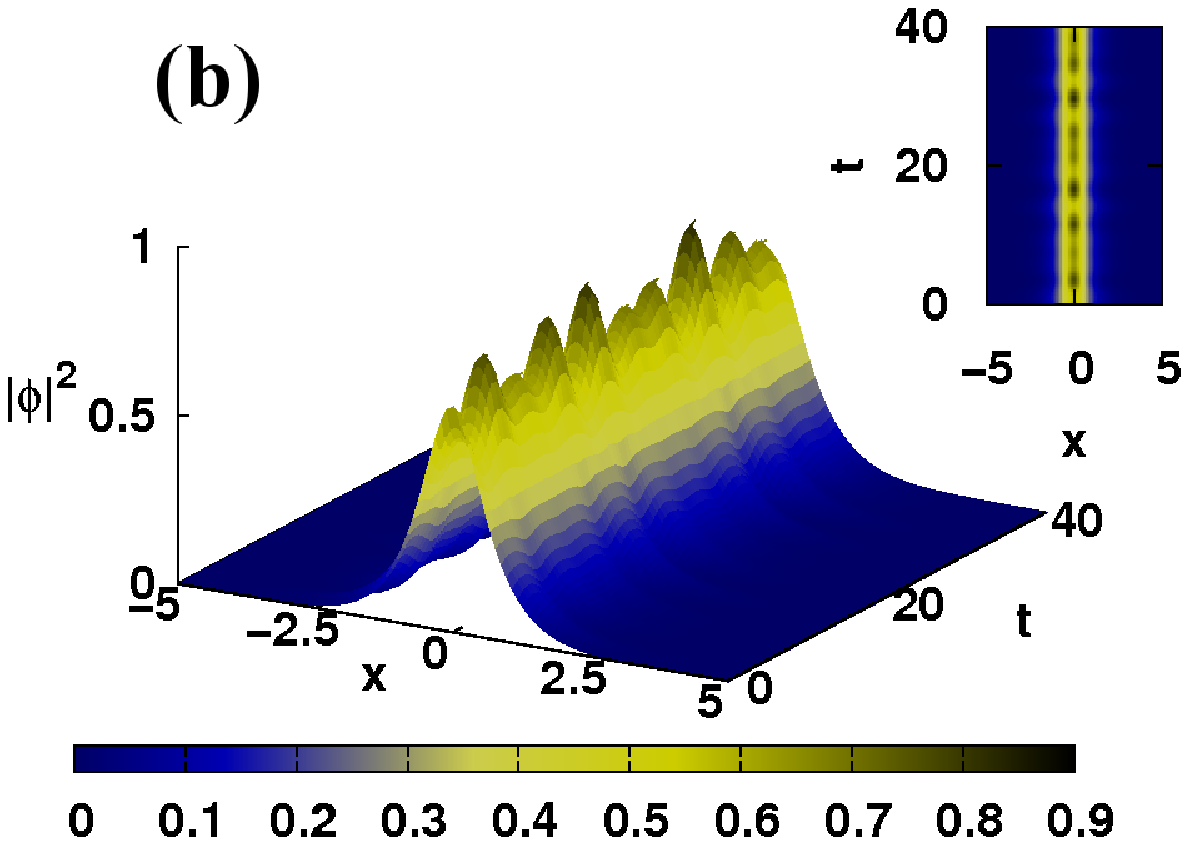}
\caption{Plot of $|\protect\psi_2|^{2}$ (Eq.~(\protect\ref{sol23})) for the cases (a) periodic and (b) quasiperiodic.}
\label{s3}
\end{figure}

FIG. 6: Plot of the potentials show in Fig.~\protect\ref{pot1} for $x$ small. In (a) we display the case periodic, and in (b) the quasiperiodic case is shown.

\phantom{a}

FIG. 7: Plot of $|\protect\psi_1|^{2}$ (Eq.~(\protect\ref{sol13})) for the cases (a) periodic and (b) quasiperiodic.

\phantom{a}

FIG. 8: Plot of $|\protect\psi_2|^{2}$ (Eq.~(\protect\ref{sol23})) for the cases (a) periodic and (b) quasiperiodic.

\newpage
%%%%%%%%%%%%%%%%%%%%%%%%%%%%%%%%%%%%
\phantom{a}
%%%%%%%%%%%%%%%%%%%%%%%%%%%%%%%%%%%%
\newpage
\phantom{a}
%%%%%%%%%%%%%%%%%%%%%%%%%%%%%%%%%%%%
\newpage
\phantom{a}
%%%%%%%%%%%%%%%%%%%%%%%%%%%%%%%%%%%%
\newpage
\phantom{a}
%%%%%%%%%%%%%%%%%%%%%%%%%%%%%%%%%%%%
\newpage
\phantom{a}
%%%%%%%%%%%%%%%%%%%%%%%%%%%%%%%%%%%%
\newpage
\phantom{a}
%%%%%%%%%%%%%%%%%%%%%%%%%%%%%%%%%%%%
\newpage
\phantom{a}
%%%%%%%%%%%%%%%%%%%%%%%%%%%%%%%%%%%%
\newpage
%%%%%%%%%%%%%%%%%%%%%%%%%%%%%%%%%%%%

\end{document}